\documentclass[prl,twocolumn,floatfix,showpacs ]{revtex4-1}
\UseRawInputEncoding
\usepackage{pifont}
\usepackage{amsmath}
\usepackage{amssymb}
\usepackage{graphicx}
\usepackage{dcolumn}
\usepackage{bm}
\usepackage[colorlinks=true,linkcolor=blue,anchorcolor=blue, citecolor=cyan,urlcolor=cyan]{hyperref}
\usepackage[mathlines]{lineno}
\usepackage{ulem}

\usepackage{epstopdf}
\begin{document}
\title{A route to fully-compensated ferrimagnetic metal: electric-field annihilation of the bilayer bandgap}
\author{San-Dong Guo$^{1}$}
\email{sandongyuwang@163.com}
\author{Rongyuan Bian$^{1}$}
\author{Feng-Ren Fan$^{2}$}
\author{Alessandro Stroppa$^{3}$}
\affiliation{$^{1}$School of Electronic Engineering, Xi'an University of Posts and Telecommunications, Xi'an 710121, China}
\affiliation{$^{2}$School of Physical Science and Technology, Soochow University, Suzhou 215006, China}
\affiliation{$^{3}$CNR-SPIN, Department of Physical and Chemical Sciences, University of L'Aquila, Via Vetoio, 67100 L'Aquila, Italy}
\begin{abstract}
Fully-compensated ferrimagnet has garnered widespread attention due to its zero-net total magnetic moment and  non-relativistic global spin splitting.
In general, for a fully-compensated ferrimagnet, at least one spin channel should be gapped to ensure a zero-net total magnetic moment, which would lead to a fully-compensated ferrimagnetic (FC-FIM) semiconductor or half-metal, and  appears to limit the existence of an FC-FIM metal.
Here we propose that an FC-FIM metal can be achieved by electrically closing the gap of a bilayer system. Using  two-dimensional (2D) ferromagnetic (FM) semiconductor as building block, we examine both FM and  antiferromagnetic (AFM) interlayer couplings and distinguish unipolar magnetic semiconductor (UMS) and bipolar magnetic semiconductor (BMS) monolayers. It is concluded that an electric field can annihilate the bilayer gap and realize the FC-FIM metal only when the interlayer coupling is AFM and the building block is a UMS.  Our scheme for realizing an FC-FIM metal can be generalized to electrically tuned 2D spin-degenerate  metal with spin-layer locking. Using first-principles calculations, we have validated our proposal by taking bilayer MnOF,  bilayer $\mathrm{ScI_2}$ and  monolayer $\mathrm{Hf_2S}$ as  examples.
Our work offers an alternative route to realize the originally forbidden FC-FIM metal, paving the way for further exploration of FC-FIM metal.

\end{abstract}
\maketitle
\textcolor[rgb]{0.00,0.00,1.00}{\textbf{Introduction.---}}
Ferromagnetic materials exhibit non-relativistic spin-polarized splitting in their electronic band structures due to the breaking of time-reversal symmetry ($T$), and they have been extensively utilized in spintronic devices\cite{1a}. However, devices based on ferromagnetic materials face several limitations:  they are highly susceptible to external magnetic fields; they offer relatively low information storage density;  their operational speed is comparatively slow.
Fortunately, zero-net-magnetization systems are emerging, which offer superior spintronic performance with  ultrahigh data densities, immunity to external perturbations, and femtosecond-scale writing speeds\cite{k1,k2}. The  $PT$-antiferromagnets (the joint symmetry ($PT$) of space inversion symmetry ($P$) and $T$ symmetry) are conventional zero-net-magnetization magnets, and they possess global spin degeneracy across the entire Brillouin zone (BZ), forbidding the magneto-optical response, anomalous Hall effect, and anomalous valley Hall effect\cite{zg1,zg2}.

Recently, another category of zero-net-magnetization materials, altermagnets, have garnered significant attentions in the field of magnetism\cite{k4,k5,k6,k7,k8,k9,k10}.
Altermagnets retain the real-space appearance of a collinear conventional  antiferromagnet, but  in momentum space, they inherit and extend the essence of ferromagnetism. Without the help of spin-orbit coupling (SOC),  they can  exhibit  momentum-dependent spin splitting, with symmetries classified by $d$-, $g$-, or $i$-wave representations. Consequently, altermagnets have been shown to host  non-relativistic lifting of Kramers degeneracy, anomalous Hall and Nernst responses, spin-polarized charge currents, and the magneto-optical Kerr effect (MOKE)\cite{zg1}.

Fully-compensated ferrimagnets constitute another category of zero-net-magnetization materials as a distinct branch of ferrimagnetism\cite{f1,f2,f3,f4,f5,f6,f7,f8,f9}.
Unlike  $PT$-antiferromagnet and altermagnet, where the two spin sublattices are connected by either $P$ or rotational/mirror ($C/M$) symmetry, the two spin sublattices  in fully-compensated ferrimagnets are not connected by any symmetry, which leads to that they exhibit a momentum-independent, $s$-wave spin splitting across the entire BZ. Like altermagnets, fully-compensated ferrimagnets can also give rise to a range of phenomena, including the anomalous Hall and Nernst effects, non-relativistic spin-polarized currents, and  MOKE\cite{f4,f9-1,f9-2,f9-3,f9-4}.
Fully-compensated ferrimagnets can be realized by breaking the lattice symmetry that connects the two spin sublattices in $PT$-antiferromagnet and alternagnet, or by breaking the $C_2$ symmetry in spin space (The $C_2$ is the two-fold rotation perpendicular to the spin axis in spin space.)\cite{zg2,f4,qq3,qq4,qq5}.
 \begin{figure*}[t]
    \centering
    \includegraphics[width=0.85\textwidth]{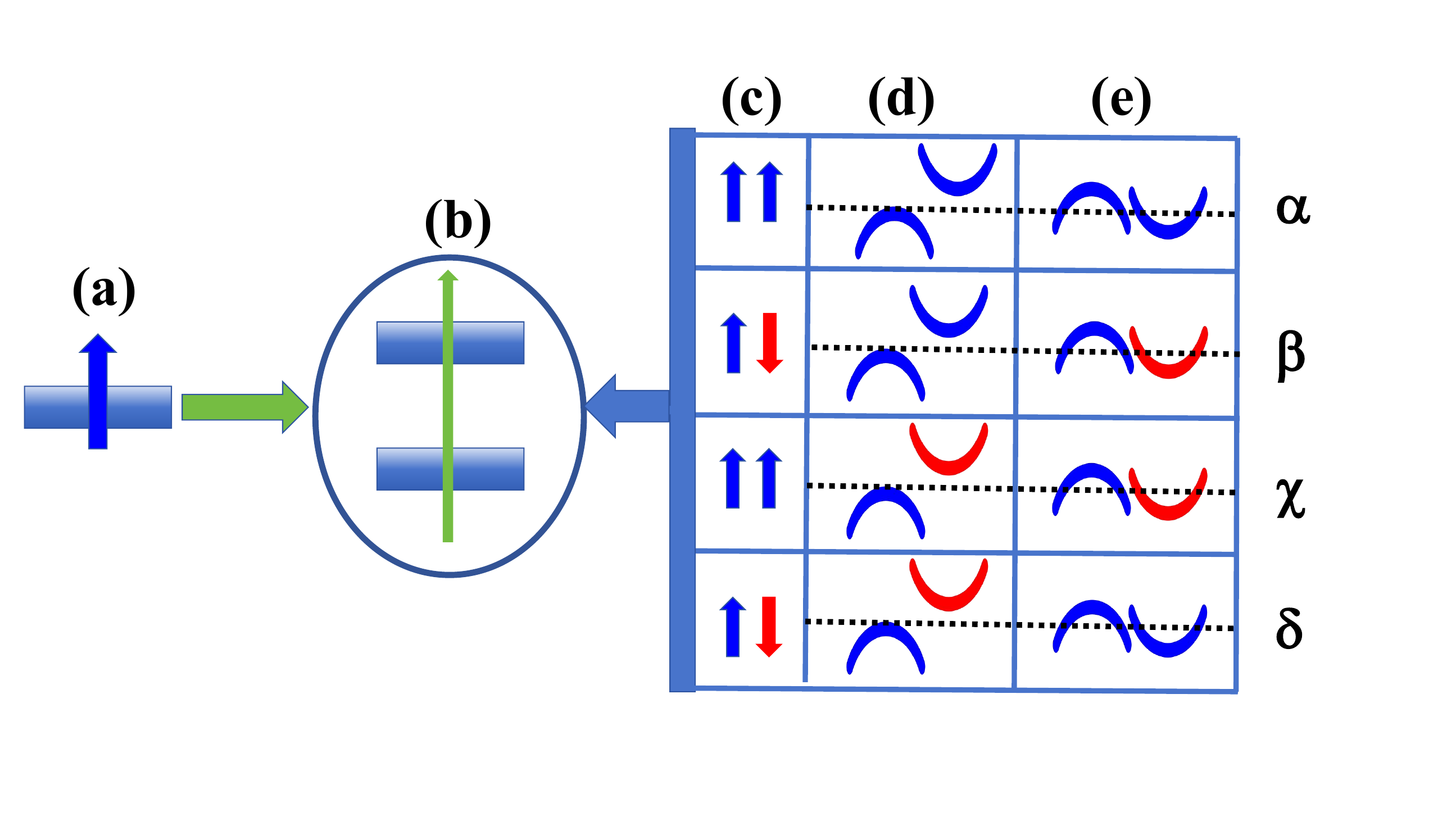}
    \caption{(Color online) (a): the fundamental building block of 2D  FM semiconductor; (b): the bilayer system under an out-of-plane electric field; (c): the interlayer coupling types, including FM ($\alpha$, $\gamma$) and AFM ($\beta$, $\delta$) coupling; (d): the band types of the building block, including unipolar ($\alpha$, $\beta$) and bipolar ($\gamma$, $\delta$) ferromagnetic semiconductors; (e): after the electric field induces gap closure in the bilayer system, its band character spans: FM half-metal ($\alpha$), FC-FIM metal ($\beta$), FM metal ($\gamma$) and FC-FIM half-metal ($\delta$).  In (a, c, d, e), the blue and red denote spin-up and spin-down, respectively. In (b), the green arrow represents an out-of-plane electric field. In (d, e), the black horizontal dashed lines indicate the Fermi level.}\label{a}
    \end{figure*}

The available studies on fully-compensated ferrimagnets have mainly focused on semiconductors or  half-metals\cite{f1,f2,f3,f4,f5,f6,f7,f8,f9}.
This is because, to ensure an exact zero-net magnetic moment, fully-compensated ferrimagnets generally require at least one spin channel to be gapped\cite{f4}.
The total number of electrons is an integer, and the number of electrons in the gapped channel is also an integer. Therefore, the number of electrons in the other spin channel must also be an integer. By appropriate electron filling, the number of spin-up and spin-down electrons can be made equal, resulting in a zero-net total magnetic moment. If both spin channels are gapped, it is a fully-compensated ferrimagnetic (FC-FIM) semiconductor; if only one spin channel is gapped, it is a FC-FIM half-metal\cite{f4}. A natural question is whether FC-FIM metal can exist.
 Here, we propose to achieve FC-FIM metal by using an electric field to modulate a bilayer system with a ferromagnetic (FM) semiconductor as building block.
 \begin{figure*}[t]
    \centering
    \includegraphics[width=0.95\textwidth]{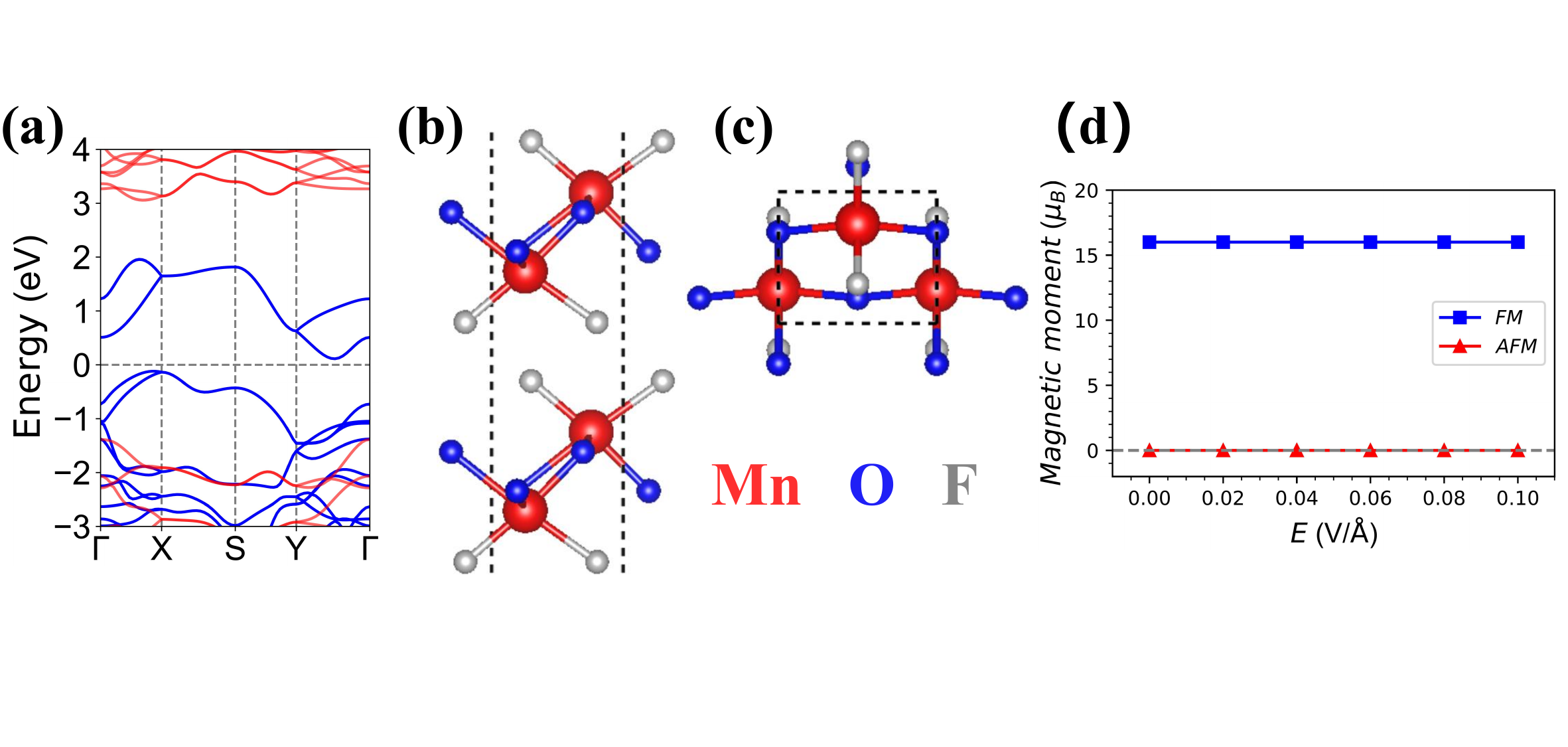}
    \caption{(Color online)(a): the energy band structures of monolayer  $\mathrm{MnOF}$;  (b, c): the  side and top views of bilayer $\mathrm{MnOF}$; (d): the total  magnetic moment per unit cell as a function of electric field $E$ for bilayer MnOF with FM and AFM interlayer coupling. In (a),  the blue and red lines denote spin-up and spin-down states, respectively.}\label{b}
\end{figure*}

\textcolor[rgb]{0.00,0.00,1.00}{\textbf{Approach.---}}
First, we construct the bilayer system using two-dimensional (2D) FM semiconductor as the basic building block  (see \autoref{a} (a, b)). Our proposal is not particularly sensitive to the lattice stacking configuration of the bilayer system.  Next, an out-of-plane electric field is applied to close the band gap of the bilayer system.
The electric field  does not directly alter the lattice structure but instead generates an electrostatic potential that affects the electronic energy states\cite{gsd1,gsd2}.
In addition, the electric field causes only a minute difference in the absolute values of the magnetic moments on the  magnetic atoms of two layers\cite{gsd1,gsd2}.
Therefore, applying an electric field to an A-type  antiferromagnetic (AFM) system can  induce spin splitting while preserving a zero-net magnetic moment, and may achieve the so-called FC-FIM metal.

The bilayer system can exhibit two types of interlayer magnetic coupling: FM and AFM cases  (see \autoref{a} (c)). The band structure of  2D FM semiconductor as the basic building block  can also be classified into two categories based on the spin character of the conduction and valence bands near the Fermi level  (see \autoref{a} (d)). In the first case, the electronic states near both the conduction band minimum (CBM) and the valence band maximum (VBM) of the FM semiconductor share the same spin character, and  such material is called unipolar magnetic semiconductor (UMS). In the second case, the states near the CBM and the VBM have opposite spin character, and this kind of material is referred to as bipolar magnetic semiconductor (BMS)\cite{s1}.

\begin{figure*}[t]
    \centering
    \includegraphics[width=0.95\textwidth]{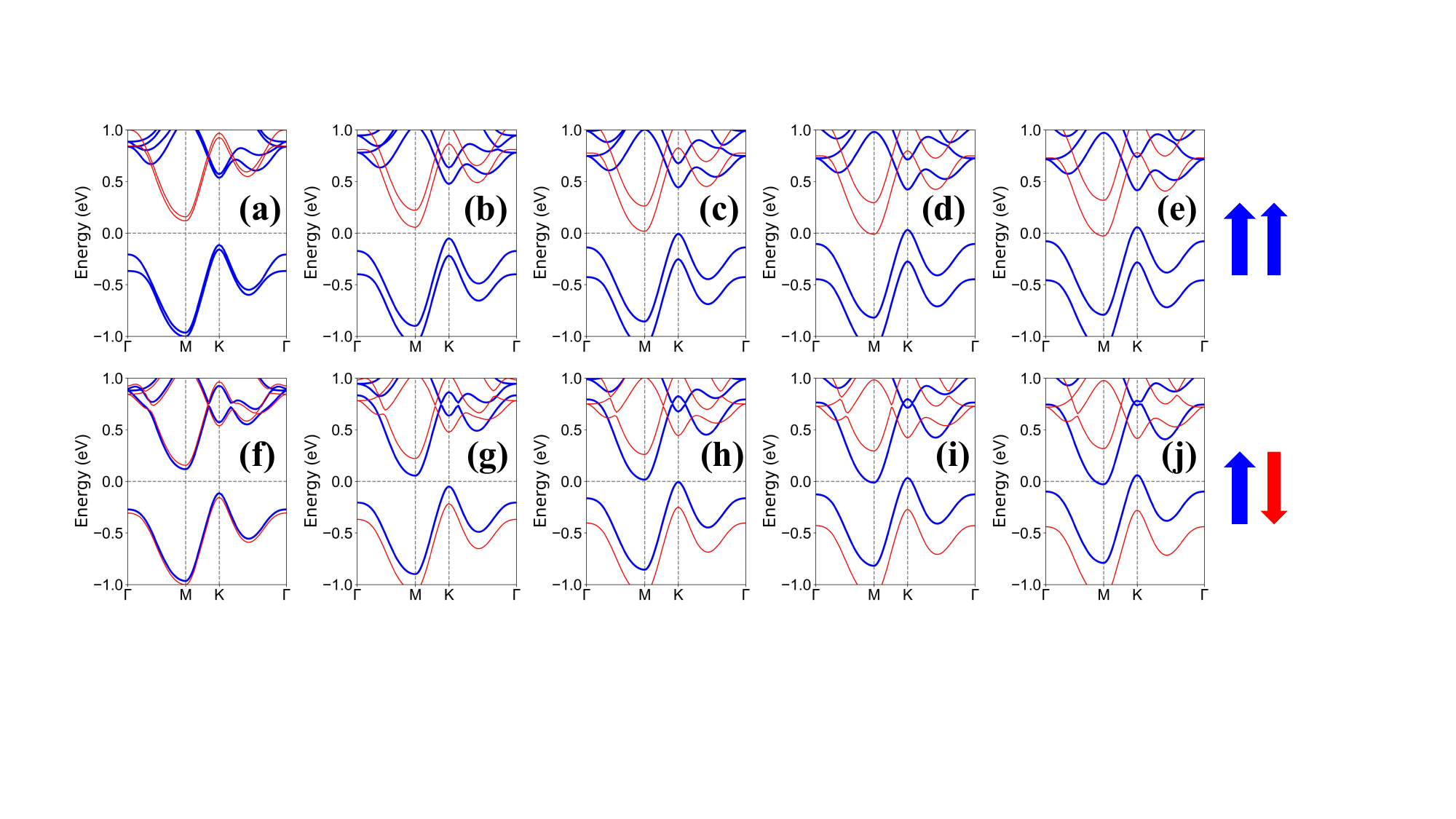}
     \caption{(Color online) For bilayer $\mathrm{ScI_2}$,  the band structures for  interlayer FM (upper panel) and AFM (lower panel) couplings under electric fields of $E$=0.00 (a, f), 0.10 (b, g), 0.20 (c, h), 0.30 (d, i), 0.40 (e, j) $\mathrm{V/{\AA}}$. The blue and red lines denote spin-up and spin-down states, respectively.}\label{e}
\end{figure*}

Because interlayer interaction is weak, we simply assume that, under FM coupling, the bands of the same spin character in the two monolayers overlap, whereas under AFM coupling, the bands of opposite spin character overlap (see FIG.S1 (a)\cite{bc}). When interlayer magnetic interaction and possible sliding ferroelectric electric field are taken into account, the band overlaps mentioned above may exhibit small splitting. Nevertheless, this does not affect our subsequent results. When an out-of-plane electric field is applied, an intuitive picture is that the electronic bands of the lower layer shift upward while those of the upper layer shift downward, leading to the closure of the band gap in the bilayer system at the critical electric field. Schematic diagram of band shifts in the lower  and upper layers under an applied electric field, including the cases of FM-UMS, AFM-UMS, FM-BMS, and AFM-BMS, are plotted in FIG.S1\cite{bc}. Depending on the type of interlayer magnetic coupling and the band character of the  building block, four distinct electronic states, including FM half-metal (\autoref{a} (e)-$\alpha$), FC-FIM metal (\autoref{a} (e)-$\beta$), FM metal (\autoref{a} (e)-$\gamma$) and FC-FIM half-metal (\autoref{e} (c)-$\delta$), can emerge in the bilayer system after the band gap closes.  Among them, FC-FIM half-metal (It is often referred to as an AFM half-metal in the literature, though strictly speaking it should be called FC-FIM half-metal from a symmetry perspective.) has been extensively studied in electrically tuned bilayer systems\cite{s2,s3,s4,fop2}. When the direction of the electric field is reversed, the order of spin splitting is also reversed.

When the interlayer coupling is AFM  and the building block is a UMS, a so-called FC-FIM metal can be realized within a suitable range of electric field after the bilayer gap closes. Below, we verify our proposal for realizing the FC-FIM metal, as well as the other three electronic states, through first-principles calculations using concrete examples.

\begin{figure*}[t]
    \centering
    \includegraphics[width=0.95\textwidth]{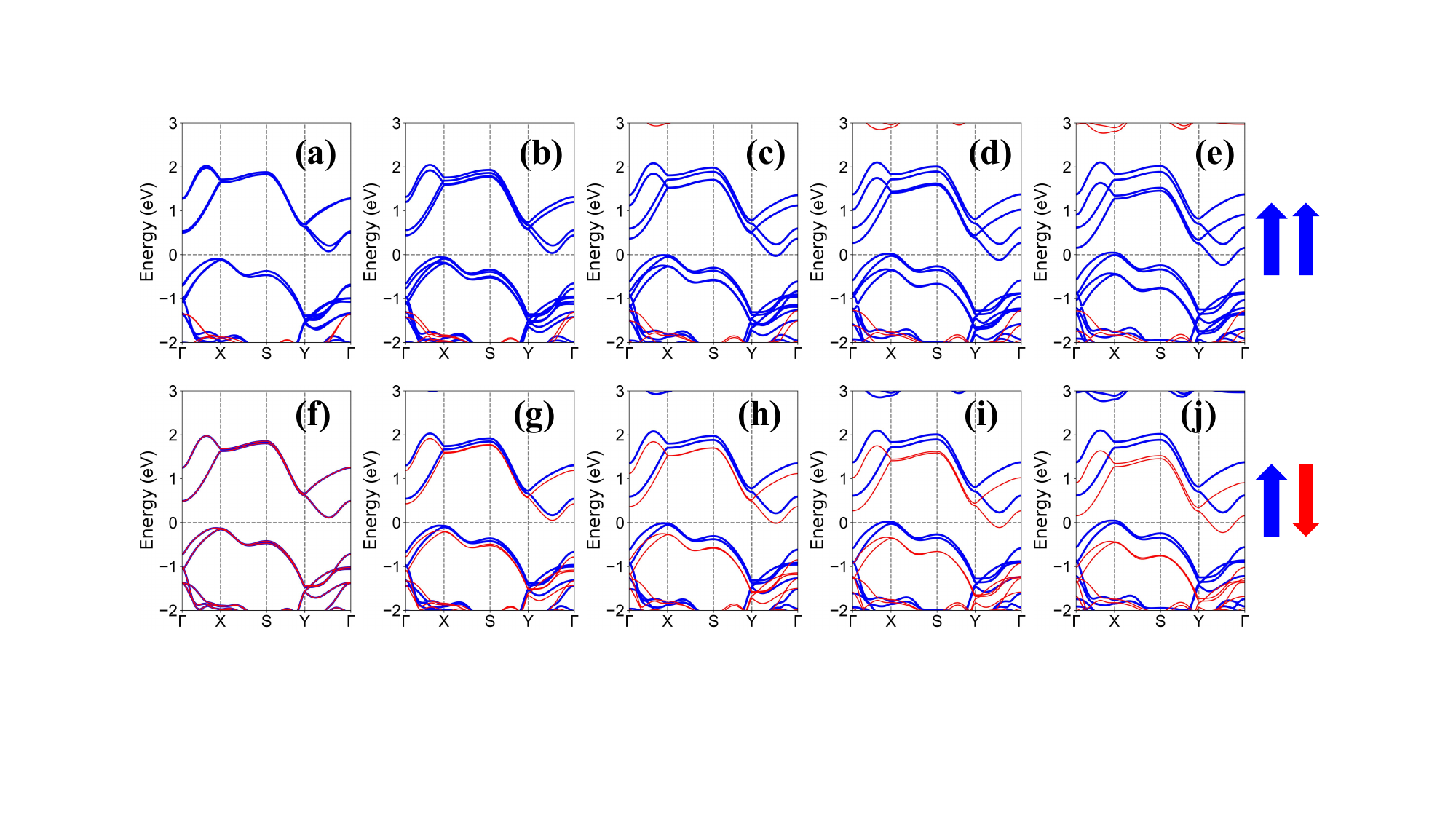}
    \caption{(Color online)For bilayer $\mathrm{MnOF}$, the  band structures for interlayer FM (upper panel) and AFM (lower panel) couplings under electric fields of $E$=0.00 (a, f), 0.02 (b, g), 0.04 (c, h), 0.06 (d, i), 0.08 (e, j) $\mathrm{V/{\AA}}$. The blue and red lines denote spin-up and spin-down states, respectively.}\label{c}
\end{figure*}

\begin{figure*}[t]
    \centering
    \includegraphics[width=0.95\textwidth]{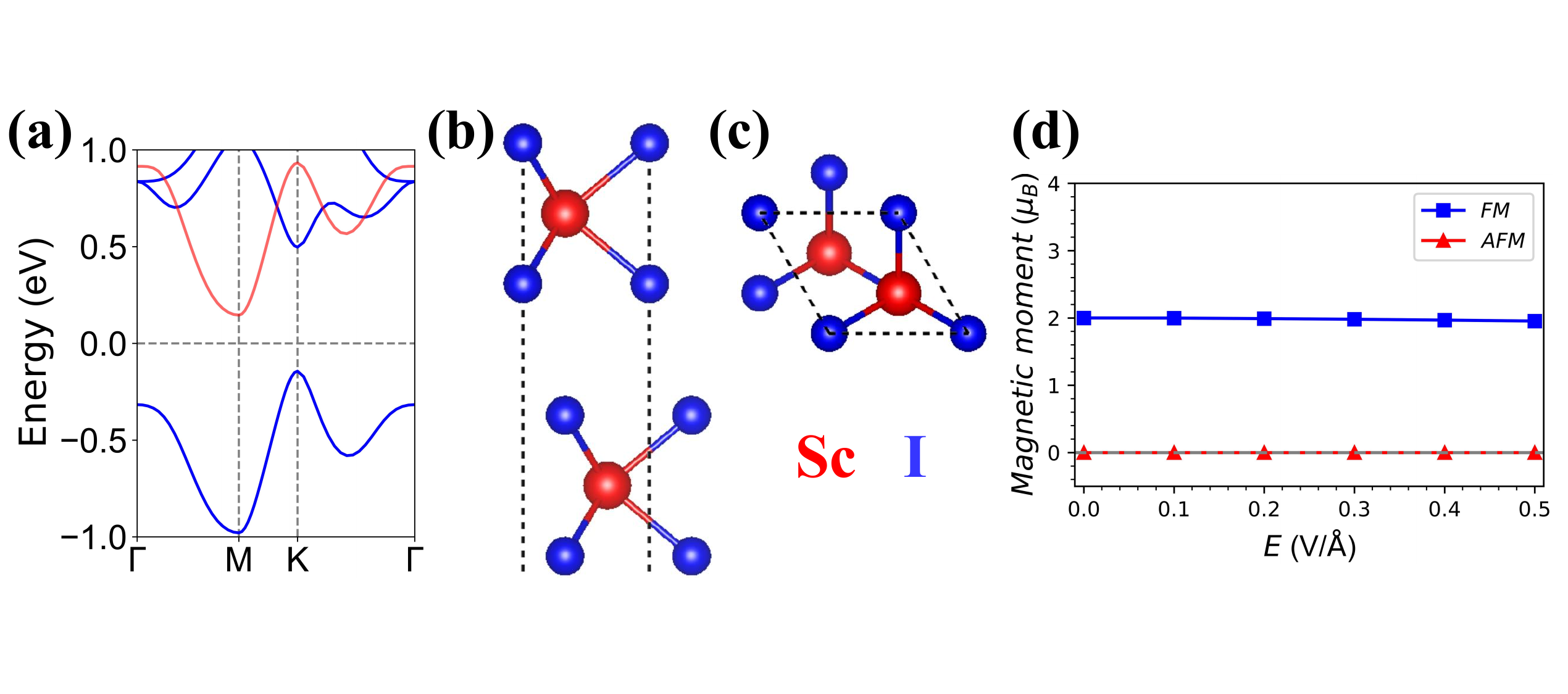}
    \caption{(Color online)(a): the energy band structures of monolayer  $\mathrm{ScI_2}$;  (b, c): the  side and top views of bilayer $\mathrm{ScI_2}$; (d): the total  magnetic moment per unit cell as a function of electric field $E$ for bilayer $\mathrm{ScI_2}$ with FM and AFM interlayer coupling. In (a),  the blue and red lines denote spin-up and spin-down states, respectively.}\label{d}
\end{figure*}

\textcolor[rgb]{0.00,0.00,1.00}{\textbf{Computational detail.---}}
The spin-polarized first-principles calculations are carried out within density functional theory (DFT)\cite{1}using the Vienna Ab Initio Simulation Package (VASP)\cite{pv1,pv2,pv3}.  The  Perdew-Burke-Ernzerhof generalized gradient approximation (PBE-GGA)\cite{pbe} is used  as the exchange-correlation functional for monolayer $\mathrm{Hf_2S}$, which has already been  adopted in previous works\cite{o1,o2,o3}. At the PBE-GGA level, we add Hubbard correction with $U$=4.00 eV for Mn-$d$-orbitals of  MnOF\cite{s5} and with $U$=2.50 eV for Sc-$d$-orbitals of   $\mathrm{ScI_2}$\cite{s6} within the
rotationally invariant approach proposed by Dudarev et al\cite{du}.   The calculations are performed with the kinetic energy cutoff  of 500 eV,  total energy  convergence criterion of  $10^{-8}$ eV, and  force convergence criterion of 0.001 $\mathrm{eV{\AA}^{-1}}$.  A vacuum layer exceeding 20 $\mathrm{{\AA}}$ along the $z$-direction is employed to eliminate spurious interactions between periodic images. The BZ is sampled with a 21$\times$21$\times$1 Monkhorst-Pack $k$-point meshes for both structural relaxation and electronic structure calculations. The dispersion-corrected DFT-D3 method\cite{dft3} is adopted to describe the van der Waals (vdW)
interactions.

\textcolor[rgb]{0.00,0.00,1.00}{\textbf{Material verification.---}}
First, we discuss the case where the basic building block is a UMS. We take MnOF\cite{s5} as the basic building block, and its structure is shown in FIG.S2\cite{bc}.
The MnOF comprises six atomic layers, with Mn
adopting a distorted octahedron coordination. The O atoms are located in the
middle two atomic layers, while the F atoms are located in the outermost atomic layers.  The structure follows the
prototype structure of monolayer CrSBr\cite{s7}.   The monolayer MnOF has been proved to be stable, and it is a FM semiconductor\cite{s5}.  The energy band structures of monolayer MnOF are plotted in \autoref{b} (a), showing it is a typical UMS.
Similar to the experimentally synthesized bilayer CrSBr\cite{s8}, we construct bilayer MnOF, whose structures are shown in  \autoref{b} (b, c).

We consider both FM and AFM interlayer couplings and apply an electric field for bilayer MnOF. The total magnetic moments of  both FM and AFM interlayer couplings as a function of electric field $E$ are shown in \autoref{b} (d), and the related energy band structures are plotted in \autoref {c}.  For the FM case, the total magnetic moment per unit cell remains an integer value of 16 $\mu_B$ throughout the considered electric-field range, consistent with the requirements of  FM semiconductor or half-metal\cite{f4}.
As the electric field increases, bilayer MnOF gradually transitions from a semiconductor to a half-metal, with the critical field being 0.04 $\mathrm{V/{\AA}}$.
This half-metallic character, where only one spin channel is metallic, is further confirmed by the spin-resolved density of states (DOS), as plotted in FIG.S3\cite{bc}.
For the AFM case, the total magnetic moment remains exactly zero within the considered electric-field range, satisfying the first requirement for  FC-FIM magnet\cite{f4}.
Due to the horizontal mirror symmetry $M_z$, the bilayer MnOF exhibits spin-degenerate bands everywhere in the absence of an electric field (Since the wave vector $k$ of the 2D material only has $k_x$ and $k_y$ components, we have $E_{\uparrow}(k)$=[$C_2$$\parallel$$M_z$]$E_{\uparrow}(k)$=$E_{\downarrow}(k)$.).
Upon application of an electric field, bilayer MnOF becomes a FC-FIM semiconductor with spin splitting due to broken $M_z$ symmetry. As the field is further increased, it turns into an FC-FIM metal  (zero-net total magnetic moment and global spin splitting) at a critical field of  0.04 $\mathrm{V/{\AA}}$,  and this metallic character can be further verified by the spin-resolved DOS (see FIG.S3\cite{bc}). Therefore, in the AFM-UMS case, the FC-FIM metal can be realized by electrically closing the bilayer bandgap.

Second, we discuss that the basic building block is a BMS.  The monolayer $\mathrm{ScI_2}$\cite{s6} is used as the basic building block, and its structure is shown in FIG.S4\cite{bc}. The Sc atom is in the middle layer, which is sandwiched by the I atoms
in the upper and lower layers, respectively. The  Sc atom is coordinated by six neighboring I atoms in
a trigonal prismatic geometry, which is similar the case of monolayer $\mathrm{MoS_2}$.
The monolayer $\mathrm{ScI_2}$ possesses excellent stability, and it has been predicted to an FM semiconductor\cite{s6}.  The energy band structures of monolayer $\mathrm{ScI_2}$  are shown  in \autoref{d} (a), showing it is a typical BMS.
The AB-stacked bilayer $\mathrm{ScI_2}$ has been proven to be the lowest-energy configuration\cite{s9}, and its structures are  shown in \autoref{d} (b, c).
\begin{figure*}[t]
    \centering
    \includegraphics[width=0.95\textwidth]{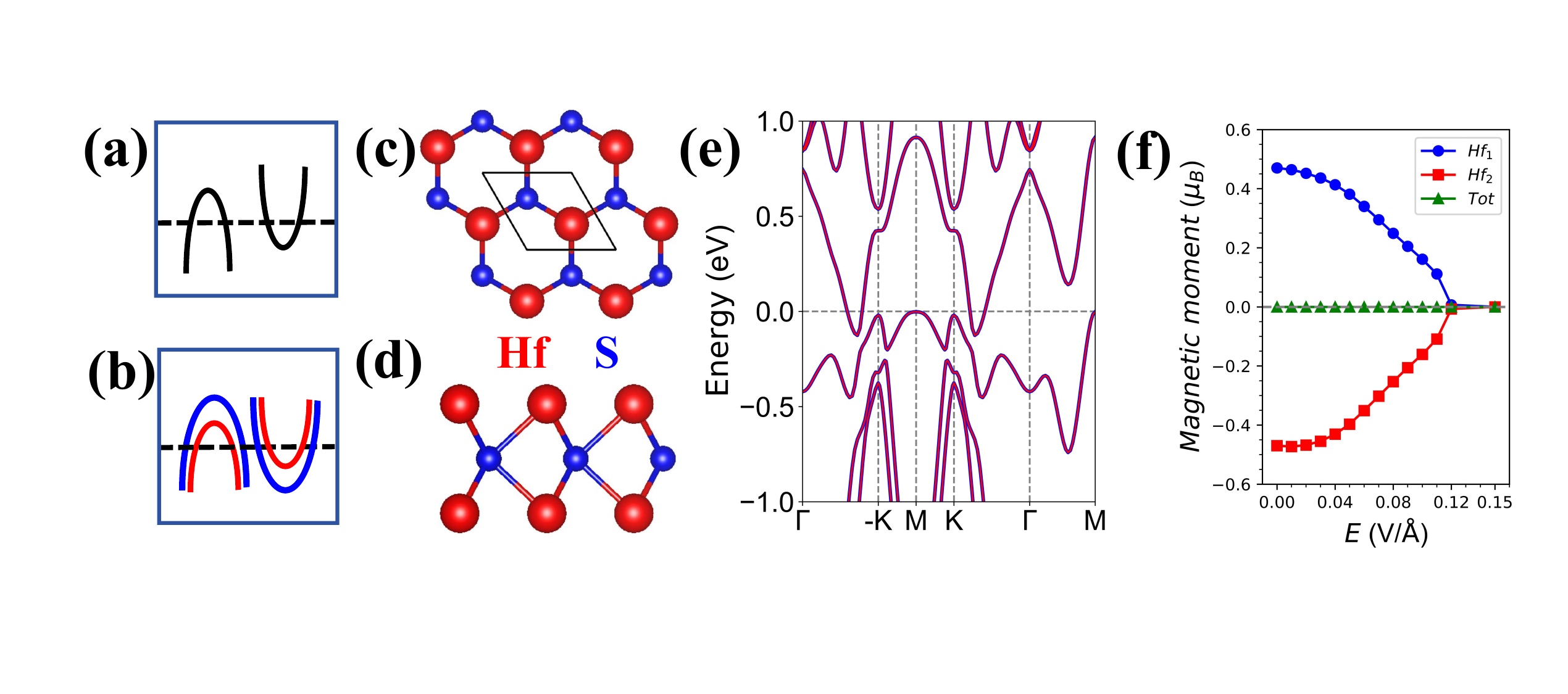}
    \caption{(Color online)(a): a 2D system with layer-spin locking  as A-type AFM ordering, exhibiting spin-degenerate band structure and being metallic;  (b): when an electric field is applied, the spin degeneracy is lifted, yet the system remains metallic; (c, d):  the top  and side  views of crystal structure of monolayer $\mathrm{Hf_2S}$; (e): the energy band structures  of  monolayer $\mathrm{Hf_2S}$; (f): the magnetic moments of the lower-layer Hf atoms (Hf$_1$), upper-layer Hf atoms (Hf$_2$), and the total magnetic  moment of the primitive unit cell ($Tot$) as functions of the electric field $E$.  In (a, b), the black, blue, and red lines denote spin-degenerate, spin-up, and spin-down states, respectively, and the black horizontal dashed lines represent the Fermi level.  In (e), the blue and red lines denote spin-up and spin-down states, respectively, and  their overlap indicates spin degeneracy. }\label{f}
\end{figure*}

We examine both FM and AFM interlayer couplings in bilayer $\mathrm{ScI_2}$ under an applied electric field. The total magnetic moment of both FM and AFM interlayer couplings as a function of electric field $E$ are displayed in \autoref{d}(d), and the corresponding energy band structures are presented in \autoref{e}.
In the FM case, when bilayer $\mathrm{ScI_2}$ is semiconducting, the total magnetic moment is an integer value of 2 $\mu_B$. As the electric field drives bilayer $\mathrm{ScI_2}$  into a metallic state, the moment slightly deviates from this integer.
With increasing electric field, bilayer $\mathrm{ScI_2}$ undergoes a continuous semiconductor-to-metal transition, and the critical field is about 0.25 $\mathrm{V/{\AA}}$. This metallicity is corroborated by the spin-resolved DOS shown in FIG. S5\cite{bc}.
In the AFM regime, the total magnetic moment stays rigorously zero across the entire applied field window, thereby fulfilling the primary prerequisite for  FC-FIM magnet\cite{f4}. In the absence of an electric field,  due to broken  $M_z$ and $P$ symmetries, the bilayer $\mathrm{ScI_2}$ exhibits intrinsic spin splitting ($E_{\uparrow}(k)$$\neq$[$C_2$$\parallel$$P$][$C_2$$\parallel$$T$]$E_{\uparrow}(k)$=[$C_2$$\parallel$$P$]$E_{\uparrow}(-k)$=$E_{\downarrow}(k)$  and $E_{\uparrow}(k)$$\neq$[$C_2$$\parallel$$M_z$]$E_{\uparrow}(k)$=$E_{\downarrow}(k)$), which is referred to as a sliding-ferroelectric-induced FC-FIM magnet from a simple physical picture\cite{qq5}. As the field is ramped up, the system switches to an  FC-FIM half-metal at the critical value of 0.25 $\mathrm{V/{\AA}}$, and the half-metallicity is unambiguously confirmed by the spin-resolved DOS (see FIG. S5\cite{bc}).
For systems with interlayer antiferromagnetism and BMS as the fundamental motif, electric-field-driven realization of an FC-FIM half-metal (commonly labeled an “AFM half-metal” in the literature) has been widely explored\cite{s2,s3,s4,fop2}.

\begin{figure*}[t]
    \centering
    \includegraphics[width=0.95\textwidth]{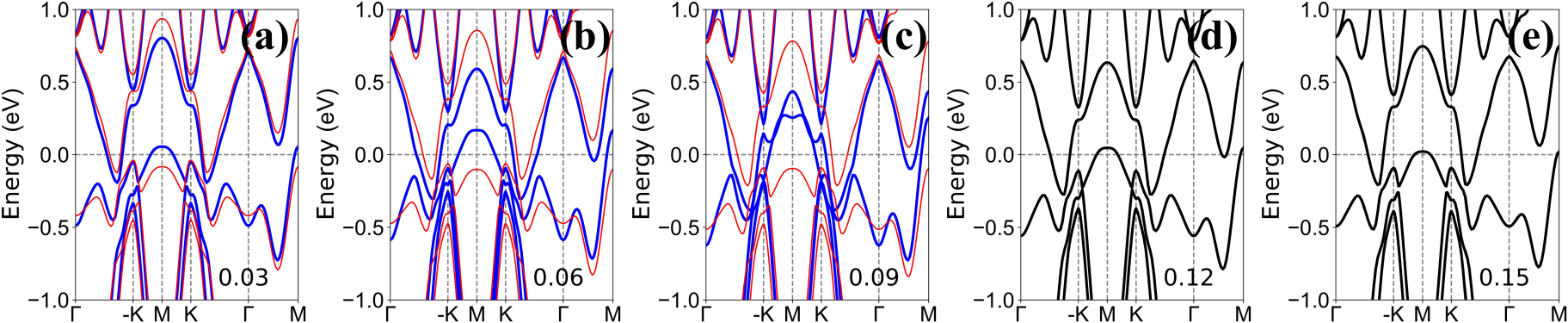}
     \caption{(Color online)For $\mathrm{Hf_2S}$,  the energy band structures  at  representative $E$=0.03 (a), 0.06 (b), 0.09 (c),  0.12 (d)  and 0.15 (e) $\mathrm{V/{\AA}}$.   In (a, b, c),  the blue (red) represents spin-up (spin-down) characters. In (d) and (e), black indicates spin degeneracy.}\label{g}
\end{figure*}

In the bilayer we have discussed, to produce the FC-FIM  metal, an electric field simultaneously opens a large spin splitting and drives the system metallic.
Consequently, our strategy can be generalized to a spin-degenerate 2D AFM metal with spin-layer locking (A-type AFM ordering), whose low-energy bands display the semi-metallic dispersion illustrated in \autoref{f}(a). Once such a host is intrinsically metallic, only an additional electric field is required to produce the spin splitting (see \autoref{f}(b)).

Monolayer $\mathrm{Hf_2S}$ is a promising candidate material, whose lattice structures are depicted in \autoref{f} (c, d). The S atomic layer of $\mathrm{Hf_2S}$  is sandwiched by two Hf atomic layers, which  possesses $M_z$ symmetry but lacks $P$ symmetry. The $\mathrm{Hf_2S}$ has been proved to be dynamically, mechanically and thermally stable, and possesses A-type AFM ordering\cite{o1,o2,o3}. The energy band structures of  $\mathrm{Hf_2S}$ are shown in \autoref{f} (e). The  symmetry [$C_2$$\parallel$$M_z$] rigidly enforces non-relativistic spin degeneracy at every $k$ point.  The  $\mathrm{Hf_2S}$ possesses  zero-net magnetic moment  and exhibits semi-metallic behavior, thereby satisfying all prerequisites for realizing  FC-FIM metal.

  The magnetic moments of the lower-layer Hf atoms (Hf$_1$), upper-layer Hf atoms (Hf$_2$), and the total magnetic moment of the primitive unit cell  as  a function of  electric field $E$ are plotted in \autoref{f} (f) . It is clearly seen that the total magnetic moment of $\mathrm{Hf_2S}$ remains zero under the applied electric field, which  ensures one of the key characteristics of  fully-compensated ferrimagnet.
 As the electric field increases, the absolute values of the magnetic moments of  Hf$_1$ and Hf$_2$ decrease, and when it reaches 0.12 $\mathrm{V/{\AA}}$, they both become zero. When the electric field is greater than or equal to 0.12 $\mathrm{V/{\AA}}$, all magnetic moments become zero, indicating that $\mathrm{Hf_2S}$ becomes non-magnetic. When the electric field is less than   0.12 $\mathrm{V/{\AA}}$, monolayer $\mathrm{Hf_2S}$  exhibits AFM ordering (see FIG.S6\cite{bc}).

The band structures representing different electric field strengths are shown in \autoref{g}.
Compared to the case without an electric field, when the electric field is less than 0.12 $\mathrm{V/{\AA}}$, the band structure exhibits a clear global spin splitting.
As can be seen from the spin-resolved density of states (see FIG.S7\cite{bc}), when an electric field is applied, the Fermi level still intersects both spin-up and spin-down states, suggesting metallic behavior.  When an electric field is applied and its strength is less than 0.12 $\mathrm{V/{\AA}}$, the total magnetic moment of $\mathrm{Hf_2S}$  is zero, there is spin splitting, and it is metallic. Therefore, $\mathrm{Hf_2S}$ with  an appropriate electric field strength  is an FC-FIM metal.

\textcolor[rgb]{0.00,0.00,1.00}{\textbf{Discussion and conclusion.---}}
The examples discussed above are primarily intended to demonstrate the feasibility of our proposal. In fact, each specific stacking configuration of a bilayer can only exhibit either FM  or AFM  interlayer coupling. Regardless of whether it's bilayer MnOF or bilayer $\mathrm{ScI_2}$, both exhibit FM interlayer coupling within the considered electric field range (see FIG.S8\cite{bc}). If future experimental techniques enable the reversal of the magnetic order in one layer, it will become possible to switch between  FM half-metal and  FC-FIM metal, or between  FM metal and  FC-FIM half-metal.
Experimentally, one needs to synthesize a narrow-gap FM semiconductor that is also a UMS, and stack it into an antiferromagnetically coupled bilayer, and then a small electric field will suffice to drive the bilayer  system into the FC-FIM metal. Very recently, the electric-field tuning of bilayer  $\mathrm{CrPS_4}$ has been demonstrated to realize a fully-compensated ferrimagnet\cite{nn}, and   an intense
electric field larger than 0.4 $\mathrm{V/{\AA}}$  has been realized  in 2D
materials by dual ionic gating\cite{zg7},  furnishing our proposal with an experimentally accessible route.

For  monolayer $\mathrm{Hf_2S}$, when SOC is included, the spin group is transformed into a magnetic group to describe its symmetry. The lattice $M_z$ symmetry will lead to: $E_{\uparrow}(k)$=$M_zT$$E_{\uparrow}(k)$=$E_{\downarrow}(-k)$. For K and -K valleys, the corresponding result is $E_{\uparrow}(K)$=$E_{\downarrow}(-K)$. The monolayer $\mathrm{Hf_2S}$ does not have lattice $P$ symmetry, so $E_{\uparrow}(k)$$\neq$$PT$$E_{\uparrow}(k)$=$E_{\downarrow}(k)$. These mean that when the SOC is taken into account, monolayer $\mathrm{Hf_2S}$ will exhibit spin splitting without an applied electric field (see FIG.S9 (a)\cite{bc}).
When an electric field is applied, the $M_z$ symmetry is also absent, which means $E_{\uparrow}(k)$$\neq$$E_{\downarrow}(-k)$. By applying an electric field, the valley polarization between the -K and K valleys will be generated since $E_{\uparrow}(K)$$\neq$$E_{\downarrow}(-K)$  (see FIG.S9\cite{bc}).

In summary,  we present  an  strategy for inducing  FC-FIM metal by electrically closing the gap of a bilayer system. Our proposal can also be extended to spin-degenerate A-type AFM metallic  monolayer. In contrast to bilayer system, here the electric field serves solely to induce spin splitting. Using first-principles calculations, we have validated our proposed scheme with specific examples. These findings provide a clear path toward exploring FC-FIM  metal.

\begin{acknowledgments}
This work is supported by Natural Science Basis Research Plan in Shaanxi Province of China  (2025JC-YBMS-008). We are grateful to Shanxi Supercomputing Center of China, and the calculations were performed on TianHe-2.
\end{acknowledgments}

\end{document}